\documentclass[final,5p,times]{elsarticle}

\usepackage{amsmath,amssymb}
\usepackage{graphicx}
\usepackage{setspace}
\usepackage{flushend}
\biboptions{sort&compress}
\usepackage{hyperref}

\begin{document}
\begin{frontmatter}
\title{On observational signatures of multi-fractional theory }

\author[label1,label2]{Mahnaz Asghari}
\ead{mahnaz.asghari@shirazu.ac.ir}
\author[label1,label2]{Ahmad Sheykhi}
\ead{asheykhi@shirazu.ac.ir}
\address[label1]{Department of Physics, College of Sciences, Shiraz University, Shiraz 71454, Iran}
\address[label2]{Biruni Observatory, College of Sciences, Shiraz University, Shiraz 71454, Iran}

\begin{abstract}
We study the multi-fractional theory with $q$-derivatives,
where the multi-fractional measure is considered to be in the
time direction. The evolution of power spectra and
also the expansion history of the universe are investigated
in the $q$-derivatives theory.
According to the matter power spectra diagrams,
the structure growth would increase in the multi-fractional model,
expressing incompatibility with low redshift measurements of
large scale structures.
Furthermore, concerning the diagrams of Hubble parameter evolution,
there is a reduction in the value of Hubble constant
which conflicts with local cosmological constraints.
Thus, primary numerical investigations imply that 
$q$-derivatives theory has no potential to relieve observational tensions. 
We also explore the multi-fractional
model with current observational data, principally Planck 2018,
weak lensing, supernovae, baryon acoustic oscillations (BAO),
and redshift-space distortions (RSD) measurements.
Numerical analysis reveals that the degeneracy between
multi-fractional parameters makes them remain unconstrained
under observations. Furthermore, observational constraints on 
$H_0$ and $\sigma_8$, detect no significant departure from 
standard model of cosmology.  
\end{abstract}
\end{frontmatter}
\section{Introduction}
In light of considerable astrophysical and cosmological observations,
it is certain that $\Lambda$CDM model affords the most appropriate
explanation of the universe. The concordance $\Lambda$CDM model described
by dark matter and also the cosmological constant $\Lambda$ as dark energy,
is introduced in the framework of general relativity (GR)
after the discovery of accelerated expansion of the universe \cite{sn1,sn2}.
The success of standard cosmological model has been also confirmed by
the majority of observational measurements including
the cosmic microwave background (CMB) anisotropies
\cite{cobe1,cobe2,wmap,cmb1,cmb2,cmb3}, large scale structures
\cite{lss1,lss2,lss3}, and baryon acoustic oscillations (BAO)
\cite{bao1,bao2,bao3}. However, despite such accomplishments,
it is known that there are some tensions between the inferred values
of cosmological parameters from local and global observational measurements.
In detail, results concerning direct determinations of Hubble constant
indicate significant discrepancies from CMB data based on $\Lambda$CDM model
\cite{H01,H02,H03,H04}.
The most recent local measurement performed by Hubble Space Telescope (HST) and the
SH0ES\footnote{Supernovae and $\mathrm{H}_0$ for the Equation of State of dark energy}
collaboration reports $H_0=73.30\pm1.04\,\mathrm{km\,s^{-1}\,Mpc^{-1}}$ \cite{H05},
which is in $5\sigma$ difference with Planck 2018 results \cite{cmb3}.
Also there is a less significant discrepancy known as $\sigma_8$ tension,
where the value of structure growth parameter $\sigma_8$ predicted from 
low redshift observations is inconsistent with Planck data \cite{s8}.
Thus, these observational tensions might imply that there is a possibility
for new physics beyond the standard $\Lambda$CDM model.
In this direction, one can consider multi-fractional scenarios which are
categorized to four independent types of theories,
mainly ordinary, weighted, $q$- and fractional derivatives \cite{cal1,cal2,cal3}.
Formerly, R. A. El-Nabulsi proposed a formalism known as the fractional
action-like variational approach to discuss the application of
fractional calculus in cosmology \cite{rami1,rami2,rami3}.
More investigations on implications of fractional approach
in GR and cosmology are also performed in literature
\cite{frc1,frc2,frc3,frc4,frc5,frc6,frc7,frc8,frc9,frc10,frc11}.
Thereafter, multi-fractional theories in which the geometry is
characterized by a fundamental scale, were proposed to improve
the physical interpretation of quantum gravity \cite{cal4,cal5,cal6}.
For some related studies on multi-fractional theories refer to
\cite{mf1,mf2,mf3,mf4,mf5,mf6,mf7}.
In the present study, we are interested in multi-fractional theory
with $q$-derivatives, which describes a more intuitive
multi-fractal spacetime \cite{q1,q2,q3,q4,q5,q6}.
We revise the observational tensions in the framework of
multi-fractional theory, and also investigate the
$q$-derivatives theory with cosmological probes.

The organization of our paper is as follows. In section \ref{sec2}
we explain modified field equations in multi-fractional spacetime.
Section \ref{sec3} contains numerical results, as well as
derived observational constraints on cosmological parameters.
We present our conclusions in section \ref{sec4}.
\section{Field equations in multi-fractional theory with $q$-derivatives} \label{sec2}
The theory with $q$-derivatives is characterized by replacing the
coordinates $x^{\mu}$ by the multi-fractional profile $q^{\mu}(x^{\mu})$
given by \cite{cal1,q6}
\begin{equation} \label{eq1}
q^{\mu}(x^{\mu})=\int{\mathrm{d}x^{(\mu)}\,v_{(\mu)}(x^{\mu})} \,,
\end{equation}
where there is no contraction on index $\mu$ in the parenthesis. Correspondingly, by using
$\frac{\partial}{\partial q^{\mu}(x^{\mu})}=\frac{1}{v_{(\mu)}(x^{\mu})}\frac{\partial}{\partial x^{(\mu)}}$,
the metric connection and the Riemann tensor in fractional frame
are defined as \cite{cal1,q6}
\begin{align}
    & {}^q\Gamma^{\rho}_{\mu\nu}=\frac{1}{2}g^{\rho\sigma}\bigg(\frac{1}{v_{(\mu)}}\partial_{(\mu)}g_{\nu\sigma}
    +\frac{1}{v_{(\nu)}}\partial_{(\nu)}g_{\mu\sigma}-\frac{1}{v_{(\sigma)}}\partial_{(\sigma)}g_{\mu\nu}\bigg) \,, \label{eq2} \\
    & {}^q R^{\rho}_{\mu\sigma\nu}=\frac{1}{v_{(\sigma)}}\partial_{(\sigma)}\,{}^q\Gamma^{\rho}_{\mu\nu}
    -\frac{1}{v_{(\nu)}}\partial_{(\nu)}\,{}^q\Gamma^{\rho}_{\mu\sigma}+{}^q\Gamma^{\tau}_{\mu\nu}{}^q\Gamma^{\rho}_{\sigma\tau}
    -{}^q\Gamma^{\tau}_{\mu\sigma}{}^q\Gamma^{\rho}_{\nu\tau} \,. \label{eq3}
\end{align}
Also, the total action in the fractional frame takes the form \cite{cal1,q6,mf6}
\begin{align} 
S=\frac{1}{16\pi G}\int{\mathrm{d}^4q(x)\,\sqrt{-g}\big({}^qR-2\Lambda\big)}+S_m \,,
\end{align}
where the action measure is \cite{q2,mf6}
\begin{equation}
\mathrm{d}^4q(x)=\Pi_{\mu=0}^{3}\mathrm{d}q^{\mu}(x^{\mu})=\mathrm{d}^4x\,\Pi_{\mu=0}^{3}v_{\mu}(x^{\mu})=\mathrm{d}^4x\,v(x) \,,
\end{equation}
so we have
\begin{align} \label{eq4}
S=\frac{1}{16\pi G}\int{\mathrm{d}^4x\,v(x)\,\sqrt{-g}\big({}^qR-2\Lambda\big)}+S_m \,.
\end{align}
where $v(x)=\Pi_{\mu=0}^{3}v_{\mu}(x^{\mu})=v_0(x^0)v_1(x^1)v_2(x^2)v_3(x^3)$ \cite{cal1,cal2}.

Action (\ref{eq4}) results in the following field equations
in multi-fractional cosmology \cite{cal1}
\begin{align} \label{eq5}
{}^qR_{\mu \nu}-\frac{1}{2}g_{\mu \nu}({}^qR-2\Lambda) =8\pi G \,{}^qT_{\mu \nu} \,,
\end{align}
where ${}^qR_{\mu \nu}={}^q R^{\rho}_{\mu\rho\nu}$ and ${}^qR=g^{\mu\nu}\,{}^qR_{\mu \nu}$ \cite{q6}.

\begin{figure*}[ht!]
    \includegraphics[width=9.5cm]{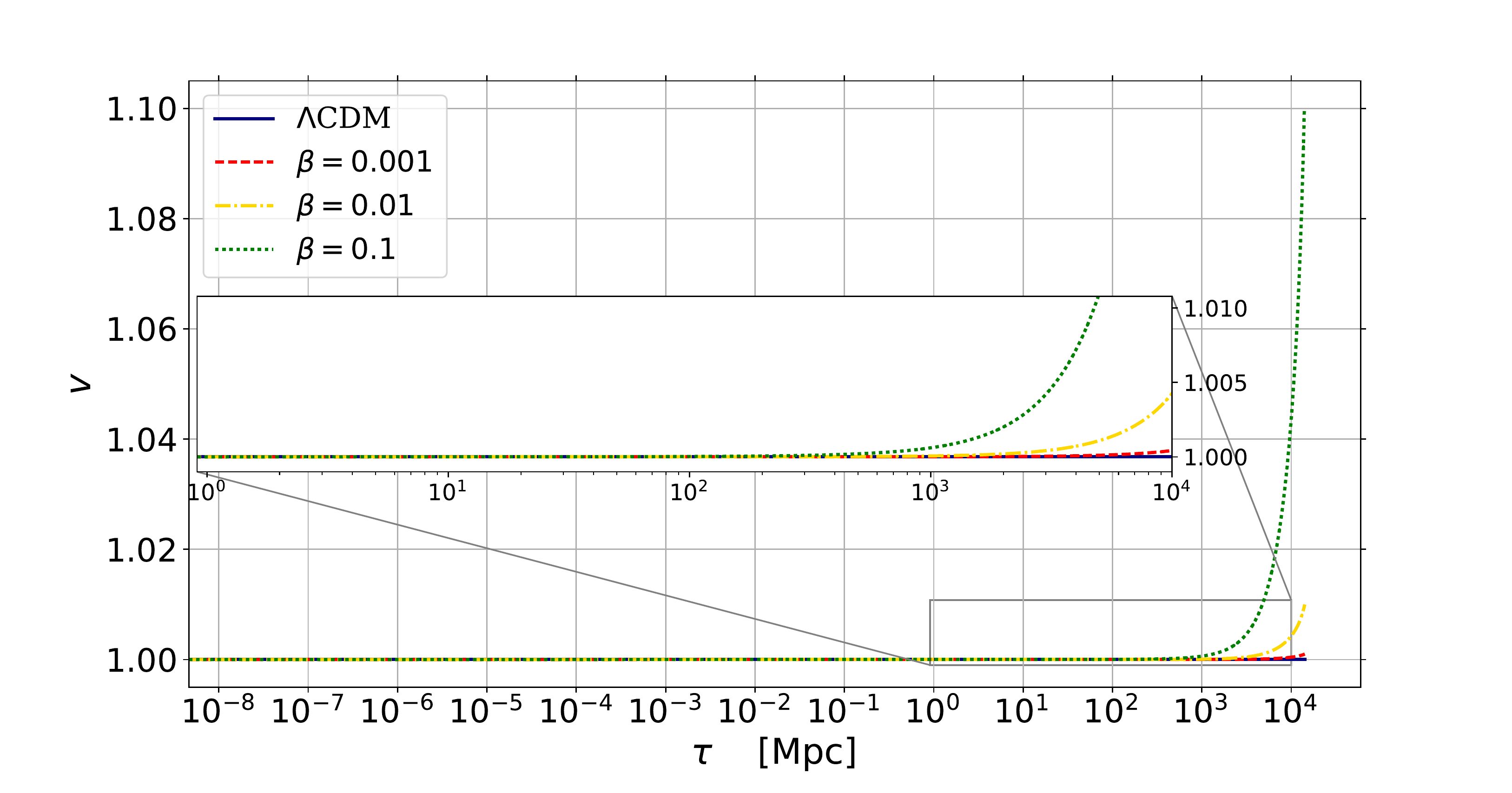}
    \includegraphics[width=9.5cm]{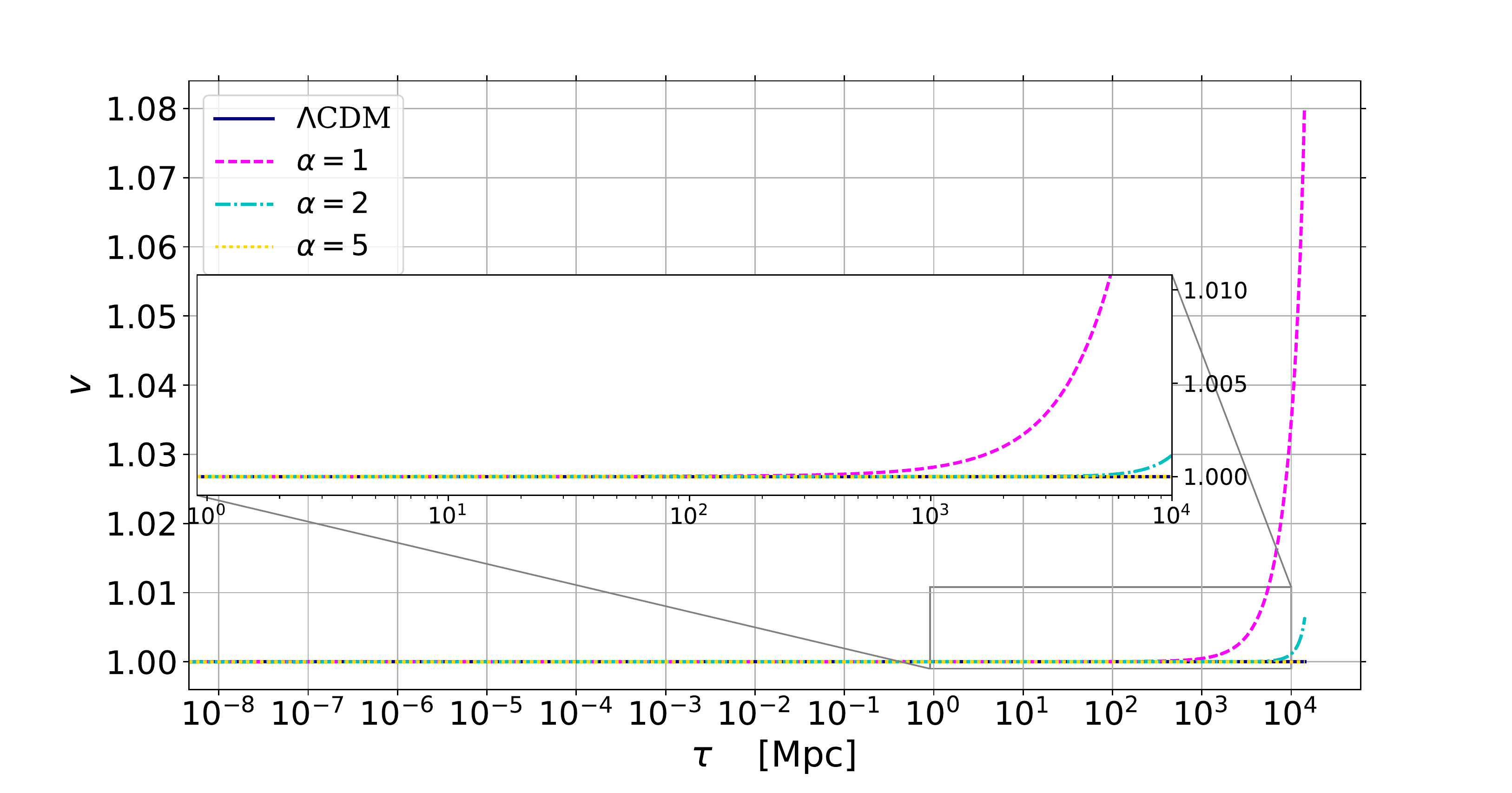}
    \caption{Left panel shows $v(\tau)$ in term of conformal time for different values of $\beta$ compared to standard model of cosmology, where $\alpha=1$, and right panel shows analogous diagrams for different values of $\alpha$, while $\beta=0.08$.}
    \label{f1}
\end{figure*}

It is more convenient to contemplate multi-fractal structure
along each direction $x^{\mu}$ as described in the literature \cite{q4},
however for the sake of simplicity, in this work we consider it
only in the time direction where $v(\tau)$ defined as
\begin{equation} \label{eq6}
v(\tau)=1+\Big(\beta\frac{a}{a_0}\Big)^{\alpha} \,,
\end{equation}
in which  $\alpha>0$ is the fractional exponent and $\beta\geq0$ is a
dimensionless constant (for more information on the range of the fractional exponent $\alpha$ refer to \cite{q6}). According to equation (\ref{eq6}) it is evident
that multi-fractional effects are more significant in late time,
as also illustrated in figure \ref{f1} (obtained from the modified version
of the CLASS\footnote{Cosmic Linear Anisotropy Solving System} code \cite{cl}).

In the following, we investigate multi-fractional theory in a
flat Friedmann-Lema\^itre-Robertson-Walker (FLRW) universe,
described by the metric
\begin{equation}
\mathrm{d}s^2=a^2(\tau)\Big(-\mathrm{d}\tau^2+\big(\delta_{ij}+h_{ij}\big)\mathrm{d}x^i\mathrm{d}x^j\Big)
\end{equation}
in the synchronous gauge, where 
\begin{equation}
h_{ij}(\vec{x},\tau)=\int \mathrm{d}^3k\,e^{i\vec{k}.\vec{x}}
\bigg(\hat{k}_i\hat{k}_jh(\vec{k},\tau)+\Big(\hat{k}_i\hat{k}_j-\frac{1}{3}\delta_{ij}\Big)6\eta(\vec{k},\tau)\bigg) \,,
\end{equation}
with $\vec{k}=k\hat{k}$,
in which $h$ and $\eta$ are scalar perturbations \cite{pt}.
Likewise, conformal Newtonian gauge is represented by the metric
\begin{equation}
\mathrm{d}s^2=a^2(\tau)\Big(-\big(1+2\Psi\big)\mathrm{d}\tau^2+\big(1-2\Phi\big)\mathrm{d}\vec{x}^2\Big) \,,
\end{equation}
with gravitational potentials $\Psi$ and $\Phi$ \cite{pt}.

We also assume the energy content of the universe as
a perfect fluid with the energy-momentum tensor
\begin{equation}
T_{\mu\nu}=\big(\rho+p\big)u_{\mu}u_{\nu}+g_{\mu \nu}p \,,
\end{equation}
which is similar to GR, since of considering
multi-fractal only in the time direction.
So, field equations in the theory with $q$-derivatives
in background level take the form
\begin{align}
& \frac{1}{v^2}H^2=\frac{8\pi G}{3}\sum_{i}\bar{\rho}_i \,, \label{eq7} \\
& \frac{1}{v^2}\bigg(2\frac{H}{a}\frac{v'}{v}-2\frac{H'}{a}-3H^2\bigg)=8\pi G \sum_{i}\bar{p}_i \,, \label{eq8}
\end{align}
in which a prime indicates a deviation with respect to
the conformal time, and $H=a'/a^2$ is the Hubble parameter. Then, from equation (\ref{eq7}),
the total density parameter for a universe consists of
radiation (R), baryons (B), dark matter (DM) and
cosmological constant ($\Lambda$) can be written as
\begin{equation} \label{eq9}
\Omega_\mathrm{tot}=\frac{1}{v^2} \,.
\end{equation}
Moreover, it is possible to write modified field equations to
linear order of perturbations in synchronous gauge (syn), given by
\begin{align}
& \frac{1}{v^2}\frac{a'}{a}h'-2k^2\eta=8\pi G a^2 \sum_{i}\delta \rho_{i(\mathrm{syn})} \,, \label{eq10} \\
& \frac{1}{v}k^2\eta'=4\pi G a^2 \sum_{i}\big(\bar{\rho}_i+\bar{p}_i\big)\theta_{i(\mathrm{syn})} \,, \label{eq11} \\
& \frac{1}{v^2}\Bigg(\frac{1}{2}h''+3\eta''+\bigg(-\frac{1}{2}\frac{v'}{v}+\frac{a'}{a}\bigg)\big(h'+6\eta'\big)\Bigg)-k^2\eta=0 \,, \label{eq12} \\
& \frac{1}{v^2}\Bigg(\bigg(-\frac{v'}{v}+2\frac{a'}{a}\bigg)h'+h''\Bigg)-2k^2\eta=-24\pi G a^2 \sum_{i}\delta p_{i(\mathrm{syn})}  \,, \label{eq13}
\end{align}
while in conformal Newtonian gauge (con) we have
\begin{align}
& \frac{3}{v^2}\Bigg(\frac{a'}{a}\Phi'+\bigg(\frac{a'}{a}\bigg)^2\Psi\Bigg)+k^2\Phi=-4\pi G a^2 \sum_{i}\delta \rho_{i(\mathrm{con})} \,, \label{eq14} \\
& \frac{1}{v}\bigg(k^2\Phi'+\frac{a'}{a}k^2\Psi\bigg)=4\pi G a^2 \sum_{i}\big(\bar{\rho}_i+\bar{p}_i\big)\theta_{i(\mathrm{con})} \,, \label{eq15} \\
& \Phi=\Psi \,, \label{eq16} \\
& \frac{1}{v^2}\Bigg[\frac{v'}{v}\bigg(-2\Psi\frac{a'}{a}-\Phi'\bigg)+\Psi\Bigg(2\frac{a''}{a}-\bigg(\frac{a'}{a}\bigg)^2\Bigg)+\frac{a'}{a}\Big(\Psi'+2\Phi'\Big)+\Phi''\Bigg] \nonumber \\
&+\frac{1}{3}k^2\Big(\Phi-\Psi\Big)=4\pi G a^2 \sum_{i}\delta p_{i(\mathrm{con})}  \,. \label{eq17}
\end{align}
Furthermore, conservation equations in the theory with
$q$-derivatives for $i$th component of the universe in background
and perturbation levels are
\begin{align}
\bar{\rho}'_i+3\frac{a'}{a}\bar{\rho}_i\big(1+w_i\big)=0 \,, \label{eq18}
\end{align}  
\begin{align}
& \delta'_{i(\mathrm{syn})}=-3\frac{a'}{a}\big(c^2_{si}-w_i\big)\delta_{i(\mathrm{syn})}-\frac{1}{2}\big(1+w_i\big)h' \nonumber \\
& \qquad\quad -\big(1+w_i\big)\bigg[\frac{9}{v}\bigg(\frac{a'}{a}\bigg)^2\big(c^2_{si}-c^2_{ai}\big)\frac{1}{k^2}+v\bigg]\theta_{i(\mathrm{syn})} \,, \label{eq19} \\
& \theta'_{i(\mathrm{syn})}=\frac{a'}{a}\bigg[3\big(w_i+c^2_{si}-c^2_{ai}\big)-1\bigg]\theta_{i(\mathrm{syn})}+v\frac{k^2c^2_{si}}{1+w_i}\delta_{i(\mathrm{syn})} \,. \label{eq20}
\end{align}
\begin{align}
& \delta'_{i(\mathrm{con})}=-3\frac{a'}{a}\big(c^2_{si}-w_i\big)\delta_{i(\mathrm{con})}+3\big(1+w_i\big)\Phi' \nonumber \\
& \qquad\quad -\big(1+w_i\big)\bigg[\frac{9}{v}\bigg(\frac{a'}{a}\bigg)^2\big(c^2_{si}-c^2_{ai}\big)\frac{1}{k^2}+v\bigg]\theta_{i(\mathrm{con})} \,, \label{eq21} \\
& \theta'_{i(\mathrm{con})}=\frac{a'}{a}\bigg[3\big(w_i+c^2_{si}-c^2_{ai}\big)-1\bigg]\theta_{i(\mathrm{con})}+v\bigg(\frac{k^2c^2_{si}}{1+w_i}\delta_{i(\mathrm{con})}+k^2\Psi\bigg) . \label{eq22}
\end{align}
Also, it is clear that choosing $v=1$ (or correspondingly $\beta=0$)
recovers field equations in standard cosmology.
It should be mentioned that in the rest of the paper,
we will investigate multi-fractional theory in the synchronous gauge.
\section{Results} \label{sec3}
This section is devoted to numerical study of the theory
with $q$-derivatives, by employing a modified version of
the CLASS code \cite{cl} according to field equations in
multi-fractional cosmology.
Furthermore, we compare multi-fractional theory with observations
by using the MCMC\footnote{Markov Chain Monte Carlo}
package M\textsc{onte} P\textsc{ython} \cite{mp1,mp2}.
\subsection{Numerical investigations} \label{sec3.1}
Here, we are interested to modify the publicly available code CLASS
to accommodate the multi-fractional field equations explained
in section \ref{sec2}, in pursuance of exploring the theory
with $q$-derivatives, numerically.
We consider Planck 2018 data \cite{cmb3} for cosmological parameters, in numerical
study.

\begin{figure*}[ht!]
    \includegraphics[width=9.5cm]{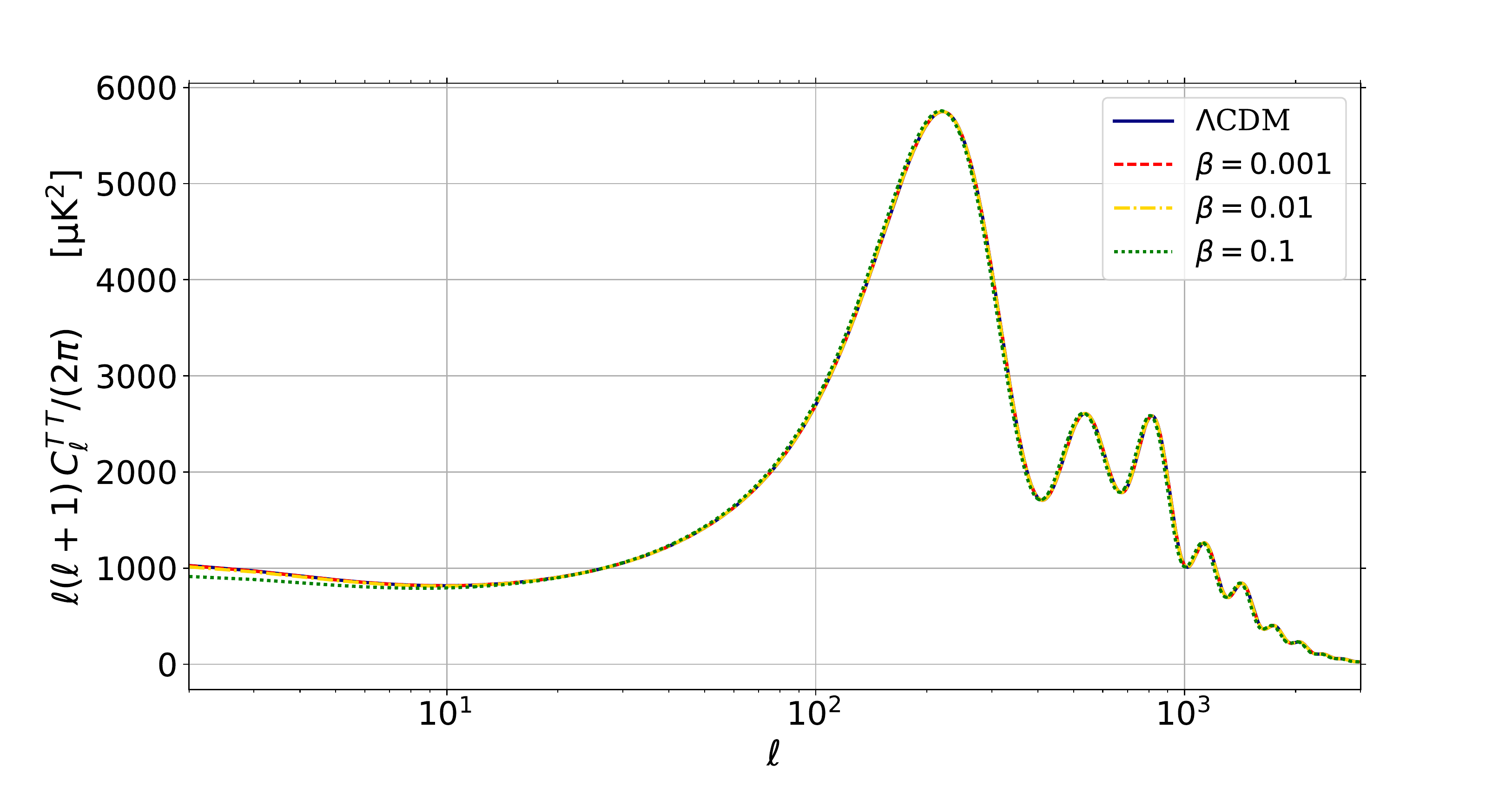}
    \includegraphics[width=9.5cm]{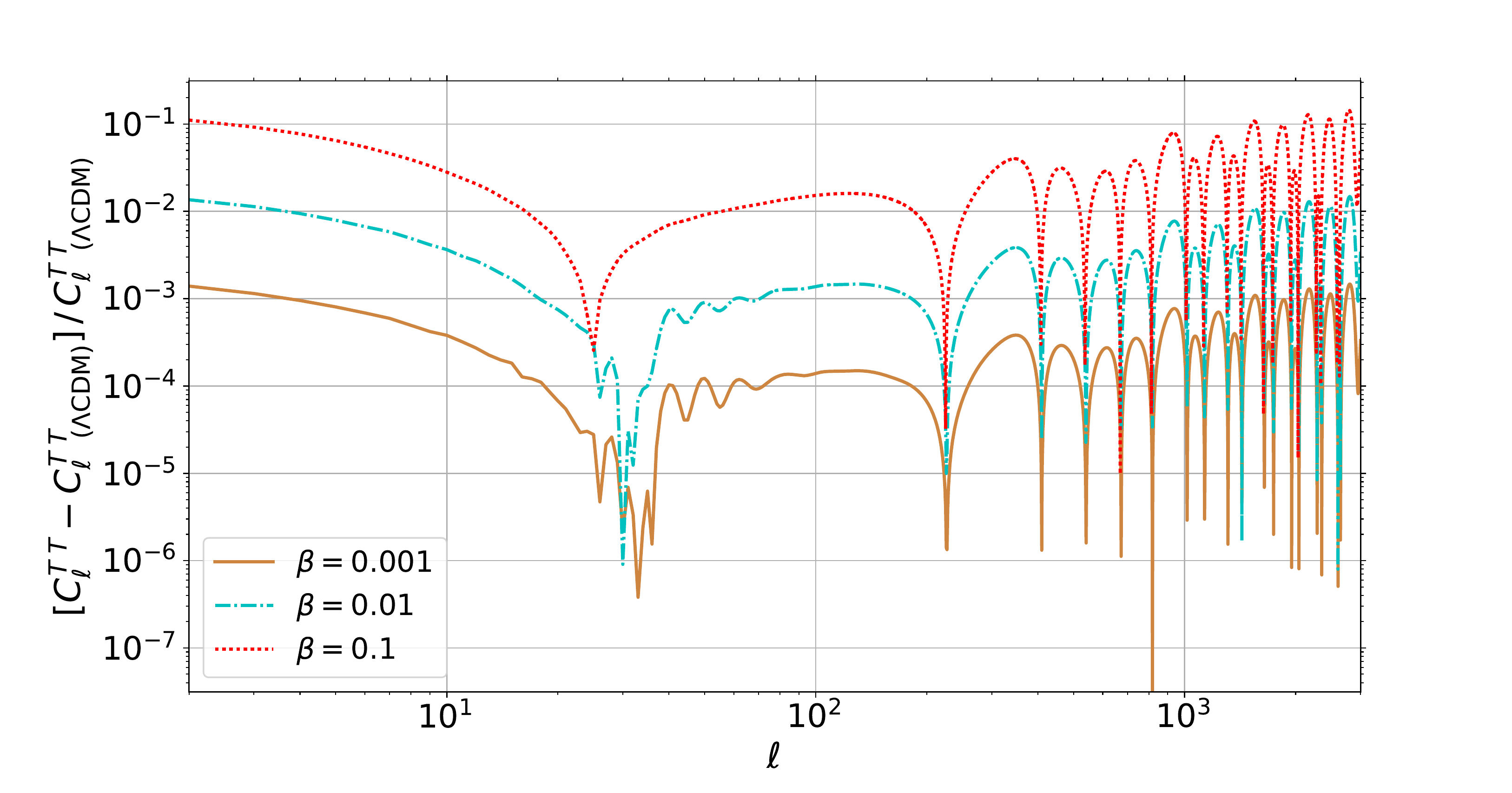}
    \includegraphics[width=9.5cm]{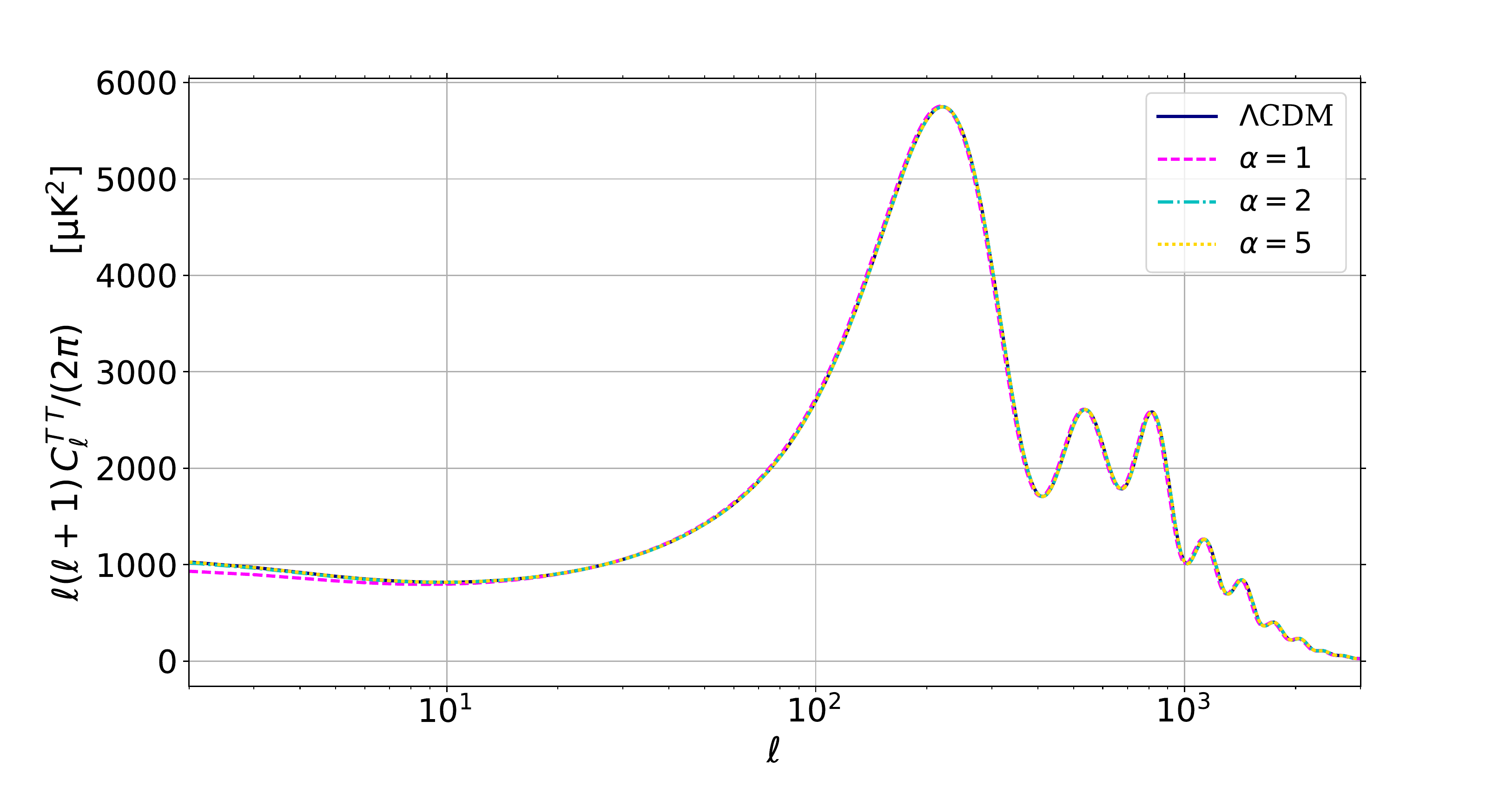}
    \includegraphics[width=9.5cm]{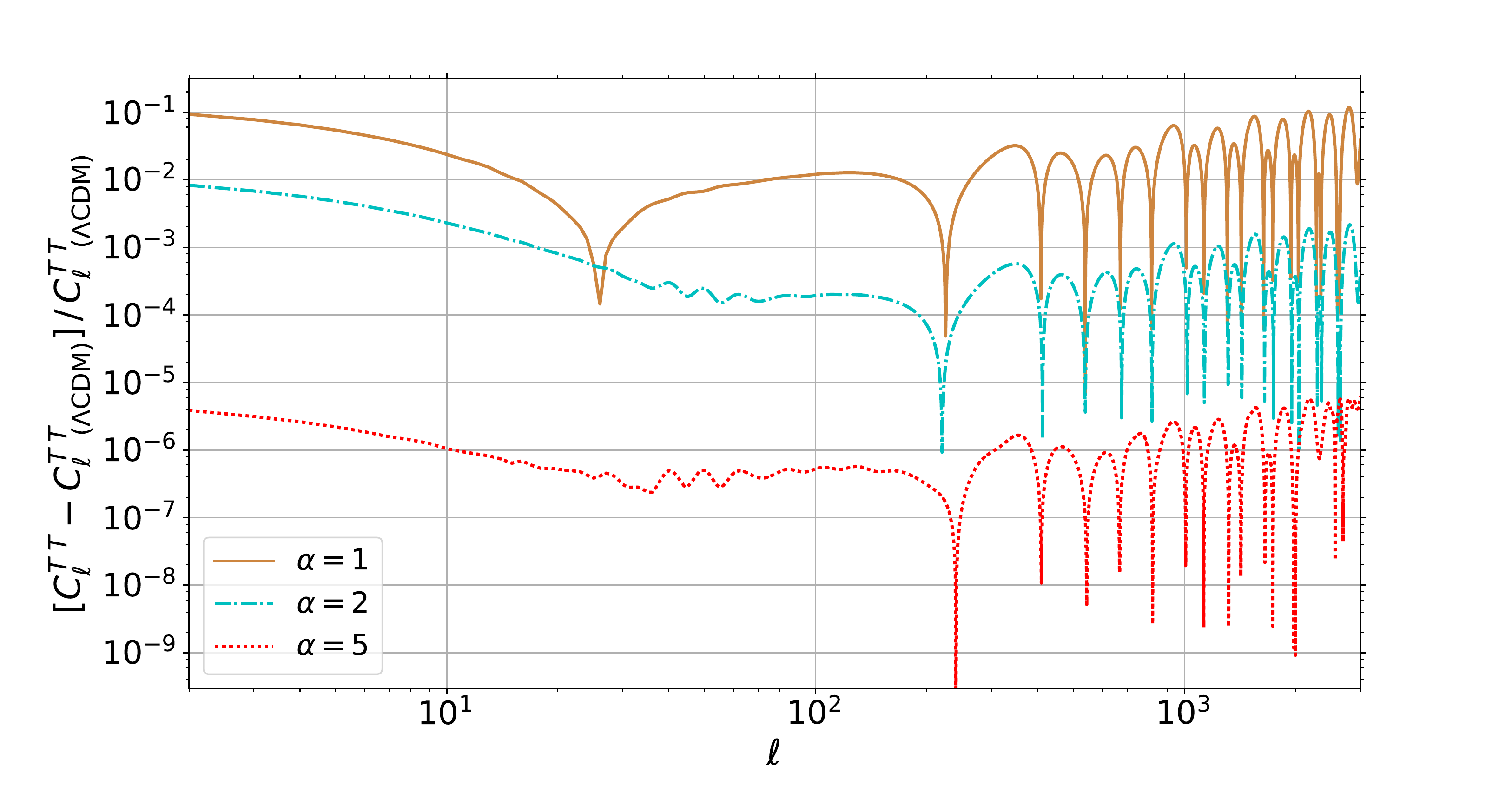}
    \caption{Upper panels show the CMB power spectra diagrams (left) and their relative ratio with respect to $\Lambda$CDM model (right) for different values of $\beta$, considering $\alpha=1$. Lower panels show analogous diagrams for different values of $\alpha$, regarding $\beta=0.08$.}
    \label{f2}
\end{figure*}

The CMB temperature anisotropy diagrams in theory with $q$-derivatives
compared to standard cosmological model, are displayed in figure (\ref{f2}).
In upper panels we can see the $TT$ component of CMB power spectra for
different values of $\beta$, while $\alpha=1$; and correspondingly,
lower panels show power spectra for different values of $\alpha$,
where $\beta=0.08$. According to this figure, it can be understood that
there is more deviation from $\Lambda$CDM model for larger values of $\beta$,
or equivalently, smaller values of $\alpha$, when the other parameter is fixed.
This feature is also predictable from modified field equations
described in section \ref{sec2}.

\begin{figure*}[ht!]
    \includegraphics[width=9.5cm]{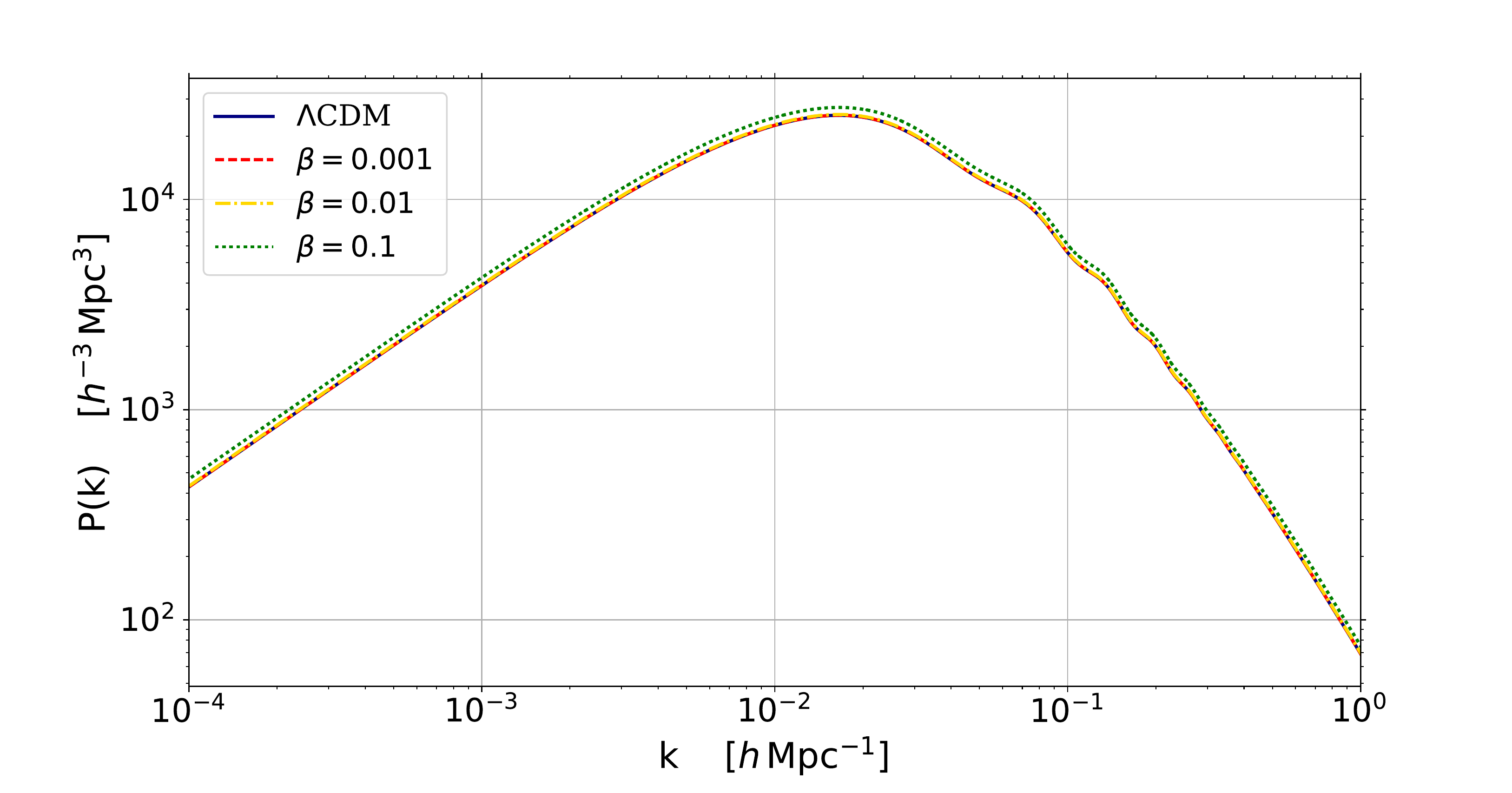}
    \includegraphics[width=9.5cm]{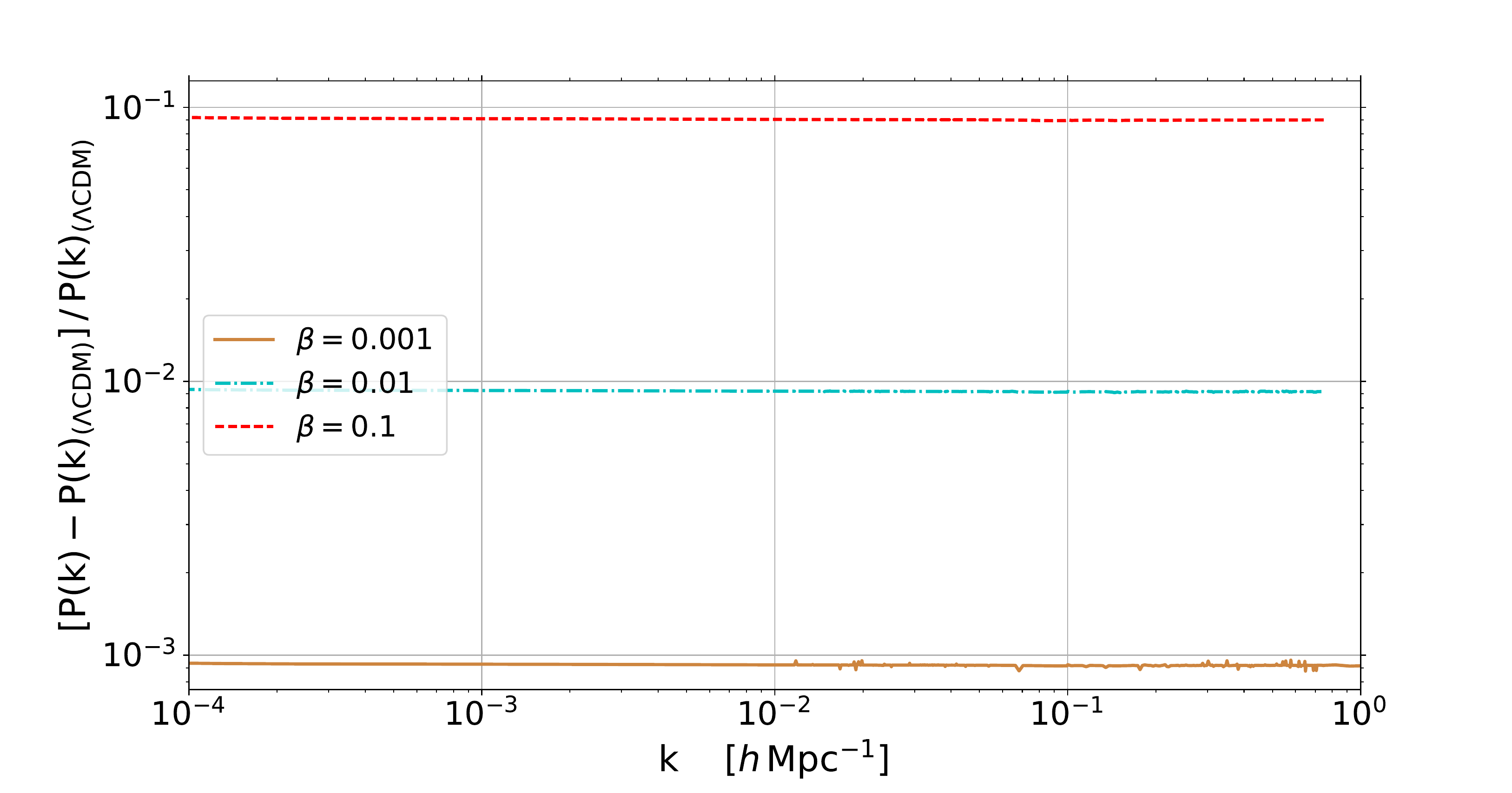}
    \includegraphics[width=9.5cm]{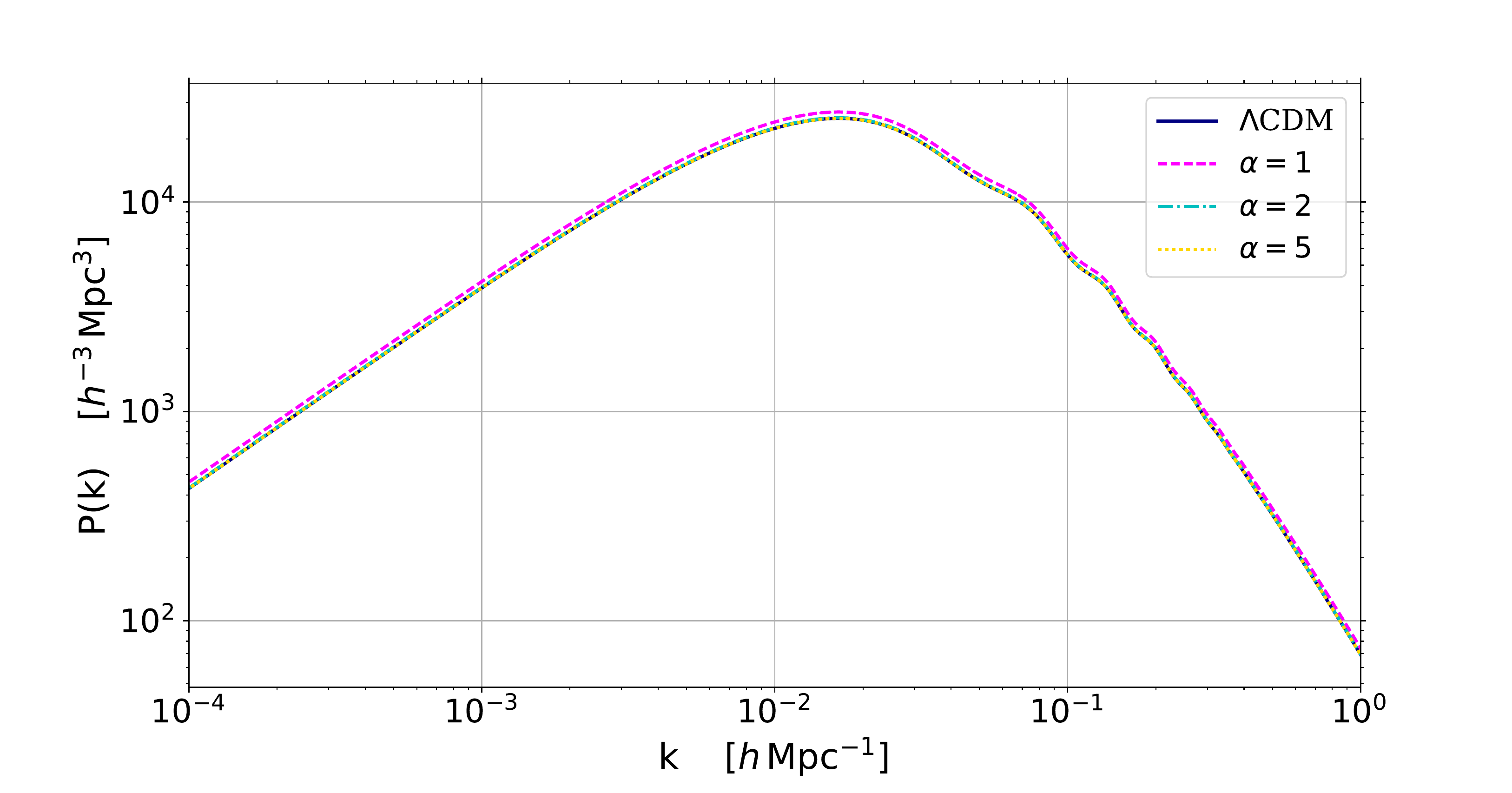}
    \includegraphics[width=9.5cm]{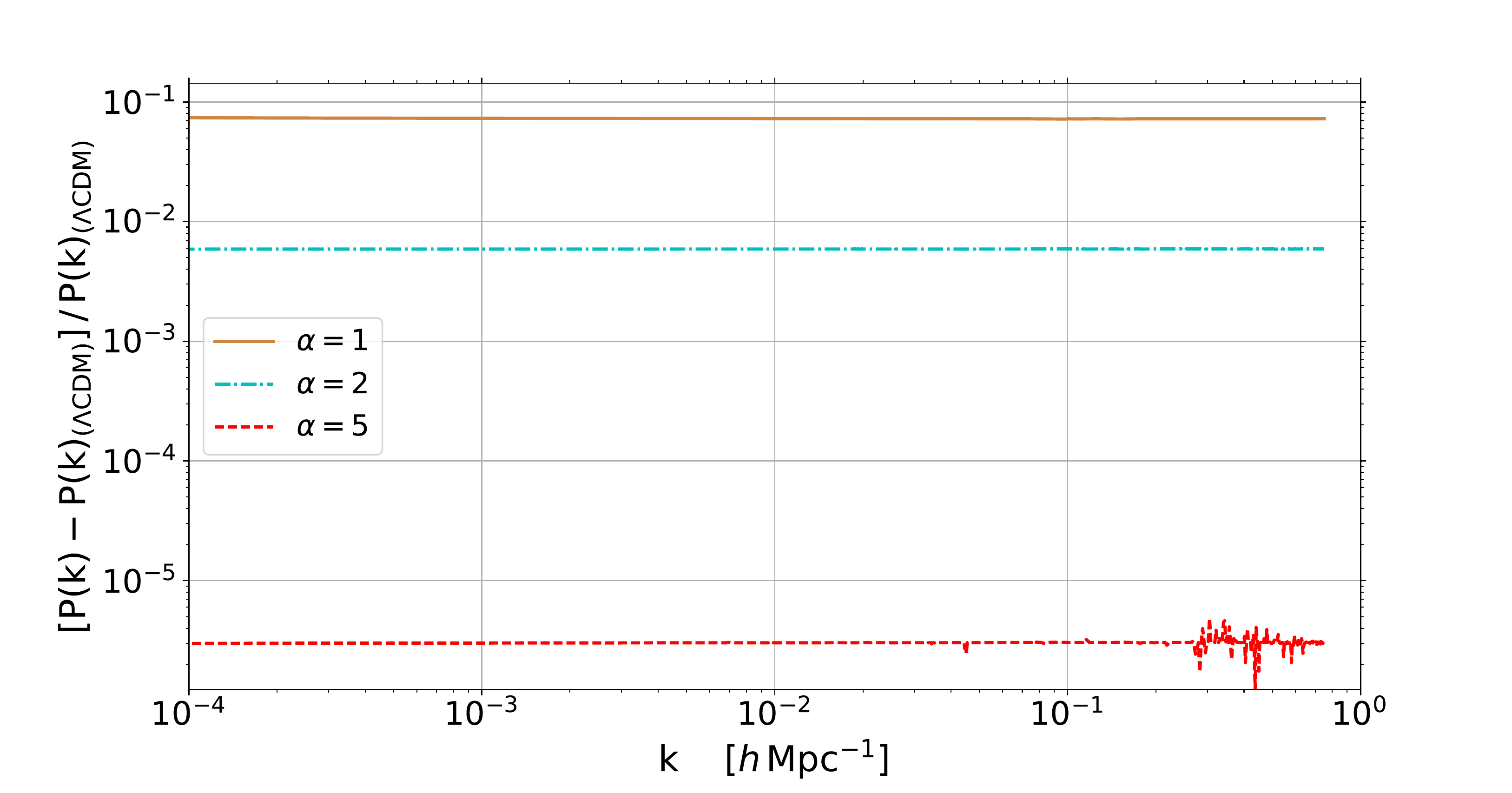}
    \caption{Upper panels show the matter power spectra diagrams (left) and their relative ratio with respect to standard cosmological model (right) for different values of $\beta$, considering $\alpha=1$. Lower panels show analogous diagrams for different values of $\alpha$, where $\beta=0.08$.}
    \label{f3}
\end{figure*}

In figure (\ref{f3}) we illustrate the matter power spectra diagrams
in multi-fractional theory comparing with $\Lambda$CDM model.
Upper panels depict power spectra for different values of $\beta$,
in which $\alpha=1$; while lower panels demonstrate power spectra
diagrams considering different values of $\alpha$, where $\beta=0.08$.
This figure expresses an enhancement in the growth of structure for
theory with $q$-derivatives, considering larger values of $\beta$,
or correspondingly, smaller values of $\alpha$
(when the other parameter is fixed),
which is in contrast with low redshift structure formation \cite{s8}.

\begin{figure*}[ht!]
    \includegraphics[width=9.5cm]{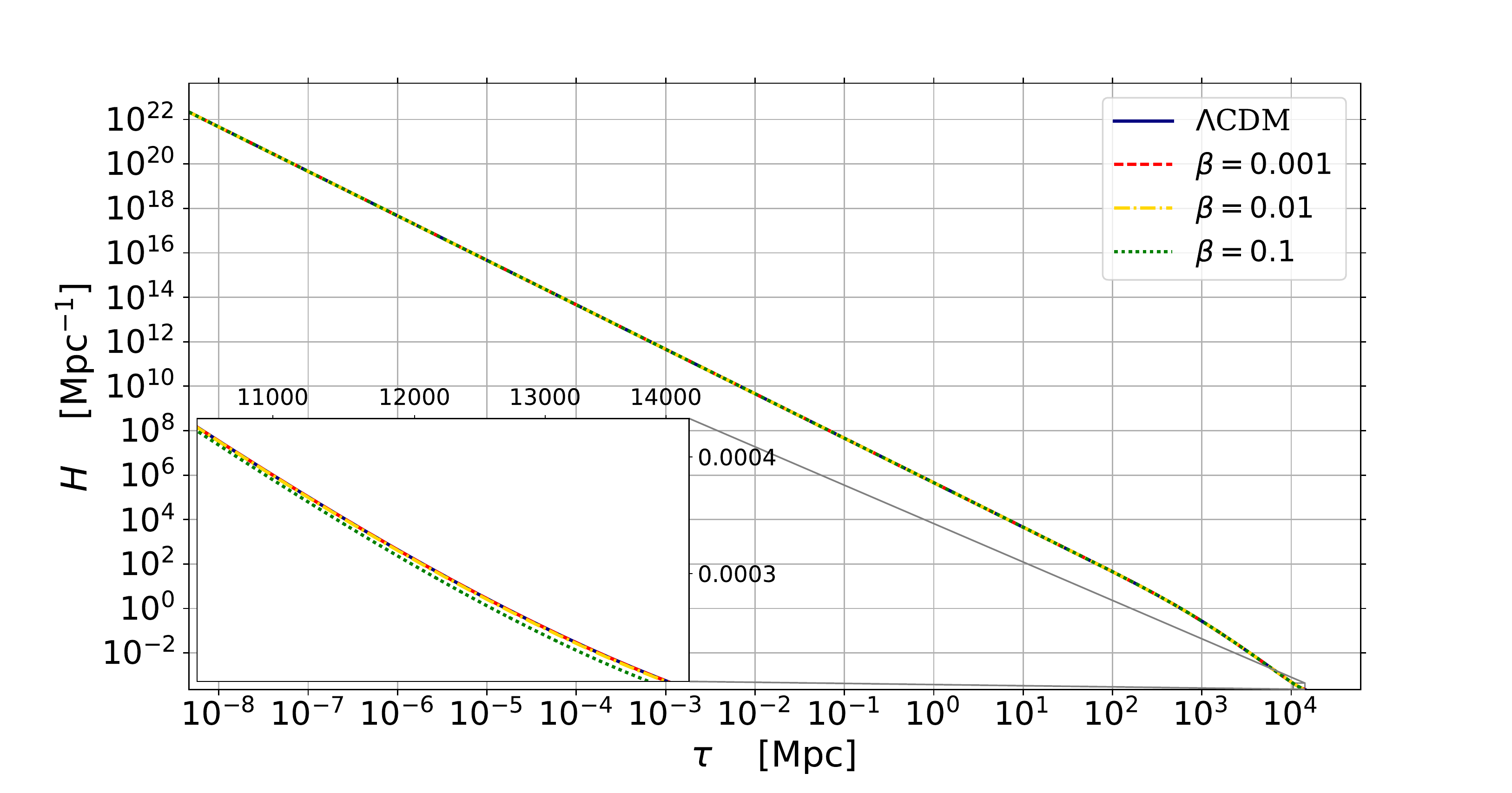}
    \includegraphics[width=9.5cm]{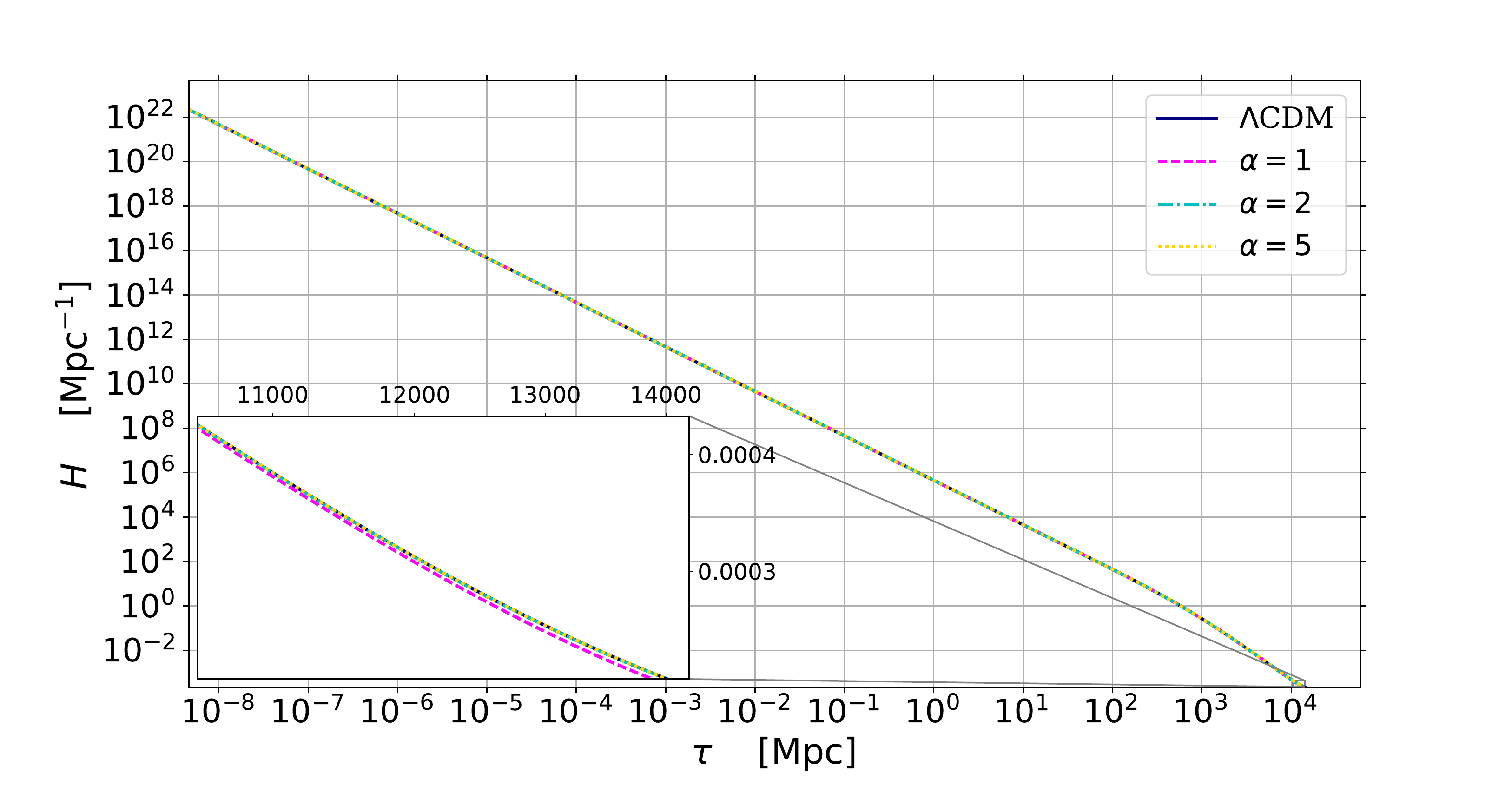}
    \caption{Left panel shows Hubble parameter in term of conformal time for different values of $\beta$ compared to $\Lambda$CDM model, while $\alpha=1$. Right panel depicts analogous diagrams for different values of $\alpha$, where $\beta=0.08$.}
    \label{f4}
\end{figure*}

Furthermore, it is exciting to contemplate the expansion history
of the universe in multi-fractional theory. Figure (\ref{f4}) depicts
the evolution of Hubble parameter in theory with $q$-derivatives,
which represents a suppression in the current value of $H$ compared to
standard cosmological model (considering larger values of $\beta$,
or equivalently, smaller values of $\alpha$). 
Thus, we can see that Hubble tension might
become more serious in multi-fractional cosmology.
\subsection{Observational constraints} \label{sec3.2}
Now we turn our attention to confronting the multi-fractional theory
with current observations by making use of the public MCMC
package M\textsc{onte} P\textsc{ython}. Accordingly,
the baseline parameter set we consider in MCMC method includes
\{$100\,\Omega_{\mathrm{B},0} h^2$,
$\Omega_{\mathrm{DM},0} h^2$, $100\,\theta_s$, $\ln (10^{10}
A_s)$, $n_s$, $\tau_{\mathrm{reio}}$, $\beta$, $\alpha$\},
where $\Omega_{\mathrm{B},0} h^2$ and $\Omega_{\mathrm{DM},0} h^2$
indicate the baryon and cold dark matter densities respectively,
$\theta_s$ is the ratio of the sound horizon to the
angular diameter distance at decoupling, $A_s$ represents the
amplitude of the primordial scalar perturbation spectrum, $n_s$
indicates the scalar spectral index, $\tau_{\mathrm{reio}}$ stands for
the optical depth to reionization,
and multi-fractional parameters $\beta$ and $\alpha$ measure
deviations from standard cosmology.
Additionally, there are four derived parameters
consist of reionization redshift ($z_\mathrm{reio}$), the matter
density parameter ($\Omega_{\mathrm{M},0}$), the Hubble constant
($H_0$), and the root-mean-square mass fluctuations on scales of 8
$h^{-1}$ Mpc ($\sigma_8$).
Also, based on preliminary numerical analysis, we choose the
prior range [$0$, $0.1$] for $\beta$, and the prior range [$1$, $5$]
for fractional exponent $\alpha$.

In order to put constraints on cosmological parameters,
we consider the following likelihoods in MCMC analysis:
the Planck likelihood with Planck 2018 data (containing high-$l$ TT,TE,EE,
low-$l$ EE, low-$l$ TT, and lensing) \cite{cmb3}, the Planck-SZ
likelihood for the Sunyaev-Zeldovich (SZ) effect measured by Planck
\cite{sz1,sz2}, the CFHTLenS likelihood with the weak lensing data
\cite{lens1,lens2}, the Pantheon likelihood with the supernovae
data \cite{pan},  the BAO likelihood with the baryon acoustic
oscillations data \cite{bao4,bao5}, and the BAORSD likelihood for
BAO and redshift-space distortions (RSD) measurements
\cite{rsd1,rsd2}.

It is worth mentioning that using different likelihoods 
to investigate cosmological models beyond the standard $\Lambda$CDM 
paradigm, can introduce unknown bias and inconsistency which 
should be considered and evaluated carefully. 
So it is important to exercise caution and test the reliability 
and suitability of these likelihoods in case of studying non-standard 
cosmological models. It is known that the Planck 2018 data is 
the most reliable and powerful probe to test cosmological models. 
Moreover, the BAO and supernovae datasets are appropriate probes 
to constrain the expansion history of the universe. 
Thus, it is convenient in the literature to apply the 
Planck likelihood combined with BAO and/or supernovae datasets 
to investigate beyond $\Lambda$CDM model. 
For some recent related investigations refer to e.g. \cite{pbs1,pbs2,pbs3,pbs4}. 
On the other hand, the Planck SZ cluster counts data, weak lensing data, 
and RSD measurements could prove useful in constraining 
the structure growth parameter $\sigma_8$. The Planck SZ and 
weak lensing measurements constrain a combination of cosmological parameters $S_8\equiv\sigma_8(\Omega_\mathrm{m}/\Omega_\mathrm{m,fiducial})^\alpha$ 
(where $\alpha$ indicates the degeneracy direction), 
in which $\Lambda$CDM is assumed as the fiducial model.
So it is expected that the obtained data to be dependent on 
the fiducial $\Lambda$CDM model. However, the cluster counts 
and weak lensing measurements usually consider some non-linear 
effects to constrain $S_8$, and since we are interested in 
linear perturbations in our study, we can assume that there is 
no need to modify the mass bias in this paper. Furthermore, 
the RSD measurements constrain the combination $f\sigma_8$, 
where $f$ is the growth rate. The combination of $f\sigma_8$ is 
independent of bias \cite{fsb}, so there is no need 
to consider modifications on bias parameter in using this likelihood. 
There are also some studies in the literature which have been 
used the combined Planck, BAO, supernovae, and RSD dataset 
to explore non-standard cosmological models \cite{pbsr1,pbsr2,pbsr3}.
Overall, while we acknowledge that using different likelihoods to 
test non-standard cosmological models can introduce bias or inconsistency, 
we believe that it is convenient to apply a combination of 
the above mentioned likelihoods in our MCMC analysis,
without considering any substantial modifications on bias parameter.
 
\begin{figure}[h!]
    \centering
    \includegraphics[width=8.5cm]{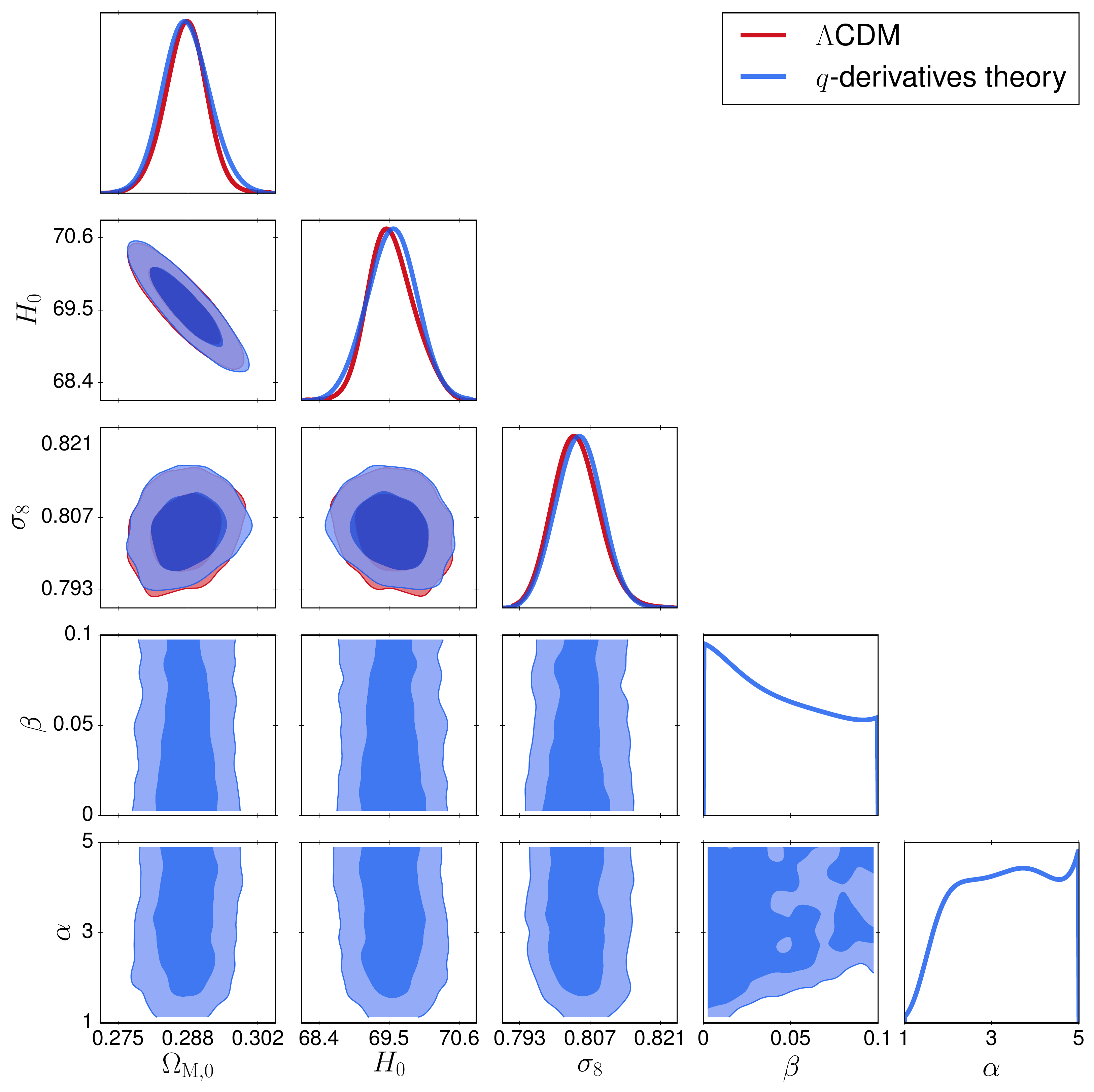}
    \caption{The $1\sigma$ and $2\sigma$ constraints on some selected cosmological parameters of multi-fractional model compared to $\Lambda$CDM.}
    \label{f5}
\end{figure}

\begin{figure*}[h!]
    \centering
    \includegraphics[width=8.5cm]{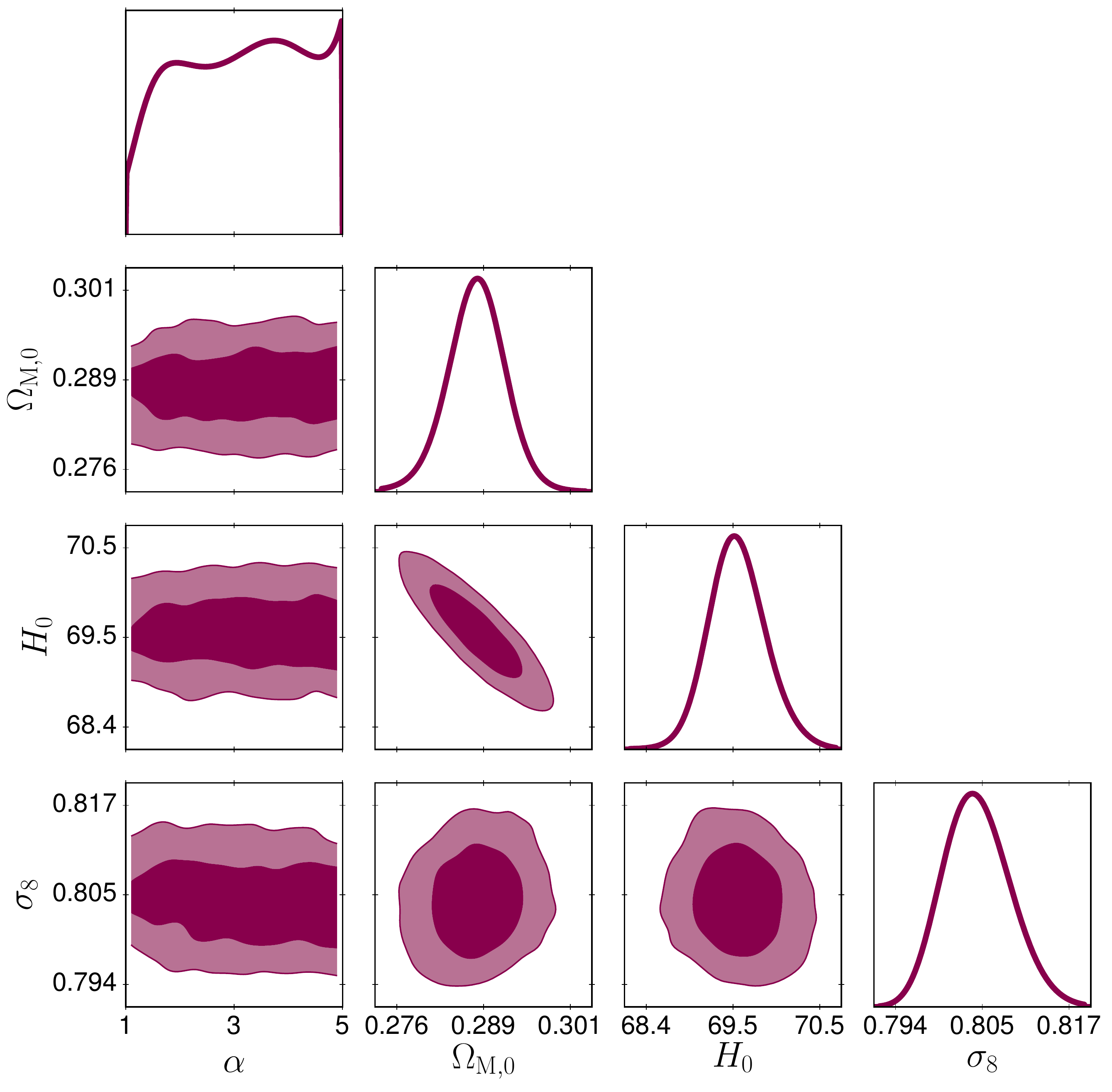}
    \includegraphics[width=8.5cm]{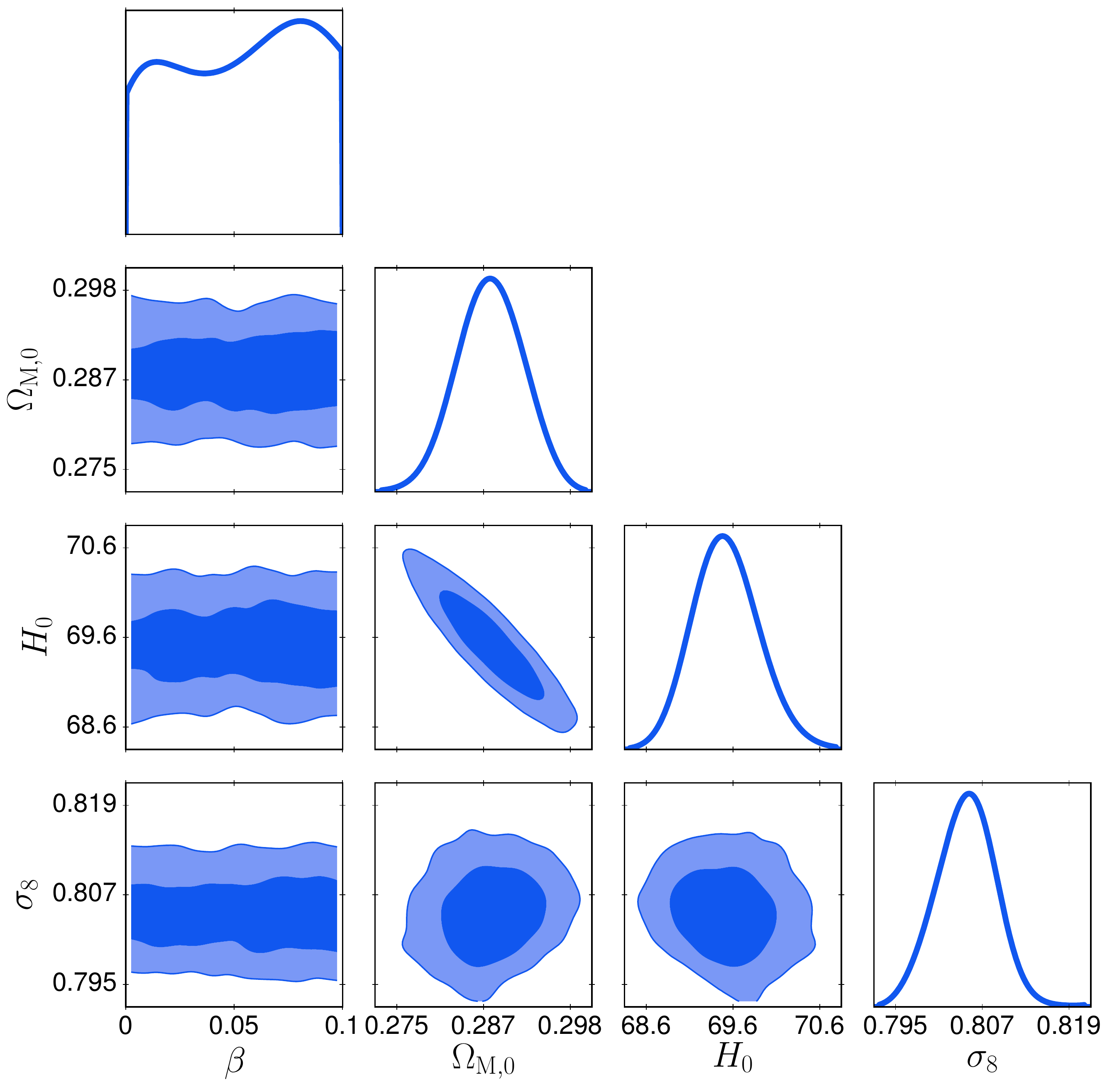}
    \caption{The $1\sigma$ and $2\sigma$ constraints on some selected cosmological parameters of multi-fractional model, while $\beta$ is fixed to $0.01$ (left panel) and $\alpha$ is fixed to $4$ (right panel).}
    \label{f6}
\end{figure*}

The report on observational constraints imposed by the
"Planck + Planck-SZ + CFHTLenS + Pantheon + BAO + BAORSD" dataset
is presented in table \ref{t1}.
\begin{table}
    \centering
    \caption{Best fit values of cosmological parameters with the $1\sigma$ and $2\sigma$ confidence levels from "Planck + Planck-SZ + CFHTLenS + Pantheon + BAO + BAORSD" dataset for $\Lambda$CDM and the theory with $q$-derivatives.}
    \scalebox{.7}{
        \begin{tabular}{|c|c|c|c|c|}
            \hline
            & \multicolumn{2}{|c|}{} & \multicolumn{2}{|c|}{} \\
            & \multicolumn{2}{|c|}{$\Lambda$CDM} & \multicolumn{2}{|c|}{$q$-derivatives theory} \\
            \cline{2-5}
            & & & & \\
            {parameter} & best fit & 68\% \& 95\% limits & best fit & 68\% \& 95\% limits \\ \hline
            & & & & \\
            $100\,\Omega_{\mathrm{B},0} h^2$ & $2.261$ & $2.263^{+0.012+0.026}_{-0.013-0.025}$ & $2.263$ & $2.264^{+0.014+0.027}_{-0.014-0.026}$ \\
            & & & & \\
            $\Omega_{\mathrm{DM},0} h^2$ & $0.1163$ & $0.1164^{+0.00078+0.0015}_{-0.00079-0.0015}$ & $0.1166$ & $0.1165^{+0.00080+0.0016}_{-0.00080-0.0016}$ \\
            & & & & \\
            $100\,\theta_s$ & $1.042$ & $1.042^{+0.00029+0.00055}_{-0.00026-0.00053}$ & $1.042$ & $1.042^{+0.00030+0.00057}_{-0.00029-0.00059}$ \\
            & & & & \\
            $\ln (10^{10} A_s)$ & $3.034$ & $3.024^{+0.010+0.023}_{-0.014-0.021}$ & $3.018$ & $3.024^{+0.0095+0.023}_{-0.013-0.021}$ \\
            & & & & \\
            $n_s$ & $0.9712$ & $0.9719^{+0.0036+0.0072}_{-0.0039-0.0074}$ & $0.9708$ & $0.9714^{+0.0038+0.0072}_{-0.0037-0.0072}$ \\
            & & & & \\
            $\tau_\mathrm{reio}$ & $0.05358$ & $0.04963^{+0.0041+0.010}_{-0.0074-0.0096}$ & $0.04628$ & $0.04933^{+0.0043+0.010}_{-0.0068-0.0093}$ \\
            & & & & \\
            $\beta$ & --- & --- & $0.01029$ & unconstrained \\
            & & & & \\
            $\alpha$ & --- & --- & $4.524$ & unconstrained \\
            & & & & \\
            $z_\mathrm{reio}$ & $7.502$ & $7.084^{+0.50+1.0}_{-0.69-1.0}$ & $6.749$ & $7.052^{+0.46+1.0}_{-0.69-1.0}$ \\
            & & & & \\
            $\Omega_{\mathrm{M},0}$ & $0.2871$ & $0.2876^{+0.0043+0.0086}_{-0.0044-0.0086}$ & $0.2879$ & $0.2879^{+0.0044+0.0090}_{-0.0046-0.0091}$ \\
            & & & & \\
            $H_0\;[\mathrm{\frac{km}{s\,Mpc}}]$ & $69.56$ & $69.54^{+0.37+0.73}_{-0.36-0.71}$ & $69.54$ & $69.52^{+0.38+0.77}_{-0.38-0.76}$ \\
            & & & & \\
            $\sigma_8$ & $0.8079$ & $0.8044^{+0.0045+0.0096}_{-0.0051-0.0091}$ & $0.8022$ & $0.8048^{+0.0041+0.0095}_{-0.0049-0.0090}$ \\
            & & & & \\
            \hline
        \end{tabular}
    }
    \label{t1}
\end{table}
The fitting results correspond to selected parameters of
multi-fractional model are also shown in figure (\ref{f5}).
Concerning numerical results, there is a degeneracy between
multi-fractional parameters $\beta$ and $\alpha$, which makes them
remain unconstrained under observational data. Accordingly,
in the direction of obtaining better constraints on these parameters,
it is recommended that we fix one of them in MCMC analysis.
Thereupon, table \ref{t2} presents corresponding results,
where we have fixed $\beta$ to $0.01$, and then $\alpha$ to $4$,
which are close to derived best fit values displayed in table \ref{t1}.
\begin{table}
    \centering
    \caption{Best fit values of cosmological parameters with the $1\sigma$ and $2\sigma$ confidence levels from "Planck + Planck-SZ + CFHTLenS + Pantheon + BAO + BAORSD" dataset for the theory with $q$-derivatives, where $\beta$ is fixed to $0.01$ in one analysis, and $\alpha$ is fixed to $4$ in the other study.}
    \scalebox{.7}{
        \begin{tabular}{|c|c|c|c|c|}
            \hline
            & & & & \\
            {parameter} & best fit & 68\% \& 95\% limits & best fit & 68\% \& 95\% limits \\ \hline
            & & & & \\
            $100\,\Omega_{\mathrm{B},0} h^2$ & $2.257$ & $2.264^{+0.014+0.026}_{-0.013-0.028}$ & $2.270$ & $2.264^{+0.014+0.026}_{-0.013-0.026}$ \\
            & & & & \\
            $\Omega_{\mathrm{DM},0} h^2$ & $0.1166$ & $0.1164^{+0.00075+0.0014}_{-0.00070-0.0015}$ & $0.1163$ & $0.1164^{+0.00082+0.0016}_{-0.00074-0.0016}$ \\
            & & & & \\
            $100\,\theta_s$ & $1.042$ & $1.042^{+0.00025+0.00056}_{-0.00028-0.00052}$ & $1.042$ & $1.042^{+0.00030+0.00060}_{-0.00029-0.00059}$ \\
            & & & & \\
            $\ln (10^{10} A_s)$ & $3.022$ & $3.025^{+0.0095+0.022}_{-0.013-0.021}$ & $3.018$ & $3.025^{+0.0093+0.022}_{-0.012-0.021}$ \\
            & & & & \\
            $n_s$ & $0.9703$ & $0.9718^{+0.0034+0.0077}_{-0.0039-0.0073}$ & $0.9701$ & $0.9718^{+0.0038+0.0068}_{-0.0034-0.0070}$ \\
            & & & & \\
            $\tau_\mathrm{reio}$ & $0.04767$ & $0.04968^{+0.0044+0.010}_{-0.0070-0.0097}$ & $0.04370$ & $0.04945^{+0.0039+0.010}_{-0.0069-0.0094}$ \\
            & & & & \\
            $\beta$ & $0.01$ (fixed) & --- & $0.07878$ & unconstrained \\
            & & & & \\
            $\alpha$ & $1.704$ & unconstrained & $4$ (fixed) & --- \\
            & & & & \\
            $z_\mathrm{reio}$ & $6.906$ & $7.089^{+0.47+0.99}_{-0.70-1.0}$ & $6.457$ & $7.064^{+0.43+1.0}_{-0.68-1.0}$ \\
            & & & & \\
            $\Omega_{\mathrm{M},0}$ & $0.2890$ & $0.2876^{+0.0043+0.0084}_{-0.0040-0.0084}$ & $0.2863$ & $0.2876^{+0.0047+0.0092}_{-0.0042-0.0090}$ \\
            & & & & \\
            $H_0\;[\mathrm{\frac{km}{s\,Mpc}}]$ & $69.40$ & $69.53^{+0.34+0.73}_{-0.37-0.71}$ & $69.68$ & $69.54^{+0.36+0.77}_{-0.40-0.75}$ \\
            & & & & \\
            $\sigma_8$ & $0.8042$ & $0.8047^{+0.0042+0.0088}_{-0.0048-0.0088}$ & $0.8008$ & $0.8046^{+0.0040+0.0088}_{-0.0045-0.0082}$ \\
            & & & & \\
            \hline
        \end{tabular}
    }
    \label{t2}
\end{table}
However, MCMC results show that multi-fractional parameters are degenerate
with other cosmological parameters, and so still remain unconstrained.
The results are more apparent in figure (\ref{f6}) which demonstrates
marginalized $1\sigma$ and $2\sigma$ confidence limit contours.
On the other hand, observational constraints on 
$H_0$ and $\sigma_8$ represent no significant deviation from 
standard cosmological model.
\section{Conclusions} \label{sec4}
This work is devoted to explore the multi-fractional theory with
$q$-derivatives by cosmological probes. Multi-fractional theories
are considered to improve the renormalization properties of
perturbative quantum gravity \cite{cal6,cal2,q5}. In
$q$-derivatives theory the coordinates $x^{\mu}$ are replaced by
the multi-fractional profile $q^{\mu}(x^{\mu})$, which results in
modified field equations described in section \ref{sec2}. We
concentrate on the multi-fractional measure in the time direction
$v(\tau)$ as defined in equation (\ref{eq6}), which is more
effective in late time. Considering numerical analysis based on
the modified version of the CLASS code, according to the
$q$-derivatives theory, it is found that observational tensions
would not be relieved in multi-fractional theory with
$q$-derivatives. Actually, matter power spectra diagrams report
larger structure formation compared to standard cosmological
model, which exhibits inconsistency with local measurements of
structure growth. On the other hand there is a suppression in the
value of Hubble constant in $q$-derivatives theory depicted in
figure (\ref{f4}), disclosing more tensions with low redshift
estimations of this parameter. Hence, according to primary numerical results, 
multi-fractional theory with $q$-derivatives is not effective in addressing 
observed tensions between low-redshift measurements and CMB data.  
To be more precise, we also perform
an MCMC calculation using CMB, weak lensing, supernovae, BAO, and
RSD data, to constrain cosmological parameters. Numerical results
indicate that because of the degeneracy between multi-fractional
parameters $\beta$ and $\alpha$ with each other and also with
other cosmological parameters, it is not possible to put
constraints on them by observational data. Moreover, 
concerning MCMC analysis, obtained constraints on $H_0$ and 
$\sigma_8$ report no considerable departure from $\Lambda$CDM model.
\section*{Declaration of competing interest}
The authors declare that they have no known competing financial interests 
or personal relationships that could have appeared to influence the work 
reported in this paper.
\section*{Data availability}
No new data were generated or analysed during the current study. 
\section*{Acknowledgments}
We thank Shiraz University Research Council.
We are also grateful to the referee for valuable comments 
which helped us improve the paper significantly.

\section*{References}
\bibliographystyle{elsarticle-num}
\interlinepenalty=10000
\bibliography{1}
\end{document}